# Haag's Theorem as a Reason to Reconsider Direct-Action Theories

R. E. Kastner

8 March 2015

ABSTRACT. It is argued that the severe consequences of Haag's inconsistency theorem for relativistic quantum field theories can be successfully evaded in the direct-action approach. Some recent favorable comments of John Wheeler, often mistakenly presumed to have abandoned his own (and Feynman's) direct-action theory, together with the remarkable immunity of direct-action quantum electrodynamics to Haag's theorem, suggest that it may well be a good time to rehabilitate direct action theories. It is also noted that, as extra dividends, direct-action QED is immune to the self-energy problem of standard gauge field QED, and can also provide a solution to the problem of gauge arbitrariness.

## 1. Introduction.

Haag's Theorem demonstrates that for interacting quantized fields, the field operators corresponding to the interacting component do not belong to the same Fock space representation as the asymptotic free fields, despite the fact that all the operators obey the same commutation relations. (See [1], §4, especially equation (57) and ensuing discussion.) Haag showed that the interacting field demands an inequivalent representation from that of the free field; the vacuum states of the two fields cannot be defined in the same representation. This result presents a serious problem for the basic mathematical consistency of quantum field theories, and has led to much discussion [2-7].]

For purposes of this paper, we can consider primarily the 'heuristic' form of Haag's Theorem, based on the notion of 'vacuum polarization.' (In [1], Haag presented a more general and formal result in which the infinite degrees of freedom of the quantized field can be seen as the actual source of the problem.) Following Earman and Fraser [2], consider a quantized scalar field ϕ with a quartic interaction represented by the Lagrangian term $\lambda \phi^4$, where $\lambda$ is a coupling constant. The heuristic version notes that the full Hamiltonian $H$ for the interacting field consists of two terms, i.e.,

$$H = H_F + H_I \qquad (1)$$

where $H_F$ is proportional to the sum of all number operators $N_k = a_k^\dagger a_k$, and $H_I$ has the form

$$H_I = \lambda \int \phi^4 \, d^3x \qquad (2)$$

Now, assuming the invariance of the vacuum state $|0_F\rangle$ of the free field under Euclidean translations, it should be the same as the vacuum state of the interacting field, $|0_I\rangle$. $|0_I\rangle$ must be annihilated by its Hamiltonian $H$. But if the 'free field' vacuum state $|0_F\rangle$ is annihilated by its Hamiltonian $H_F$, it will not be annihilated by the full Hamiltonian $H$ including $H_I$, which contains a term with a product of four creation operators not cancelled by any other contribution. (This is the 'vacuum polarization.') So we have a contradiction: $|0_F\rangle$ and $|0_I\rangle$ cannot in fact be the same state.

## 2. How direct-action theories can evade Haag's theorem

The first thing to recall is that in a direct action theory (DAT) the field interactions are not mediated by quantized fields considered as independent degrees of freedom, but instead by a direct, 'nonlocal' interaction between sources of the field. As will be discussed below, this interaction corresponds to the time-symmetric propagator of a non-quantized field theory. So from the point of view of DAT what causes the problem is the key assumption of standard QFT, namely that the interaction can be represented by field operators that create and destroy Fock space states. Drop that assumption, and we escape Haag's result, because the interacting field requires no Fock space description at all. In this picture, Fock space states describe only incoming and outgoing ('free') states. In the DAT, free states are distinguished from interacting states in that the former are those that prompt an absorber response, while the latter do not. This point will be elaborated further in Section 3.

Thus, a clean and immediate solution to the problem is to banish the notion of an independently existing field with its own degrees of freedom, and deal instead with a direct-action theory. In other words, the message of Haag's theorem is taken to be that QFT is not the correct model; a different, yet empirically equivalent, model is needed. Narlikar's work [8], as well as that of Wheeler/Feynman [10] and Davies [11-13] shows that DAT is just such a model.

While direct-action theories are widely considered to have significant drawbacks for pragmatic, computational purposes, they can be emulated to empirical equivalence by the QFT model.[1] Indeed, Narlikar [8] showed that any interacting field theory of a field $\phi$ described by the usual invariant bilinear Lagrangian of the field and its derivatives, and an interaction term $I$ of the form $I \sim g \langle \phi, j^{(i)} \rangle$, where the $j^{(i)}$ are the source currents and $g$ a coupling constant, is expressible as a direct action theory. In such a theory, the field at a point $x$ due to current $j^{(i)}(y)$ is given by

$$\phi(x) = g \int \bar{D}(x,y) j^{(i)}(y) dy \qquad (3)$$

where $\bar{D}(x,y)$ is the time-symmetric propagator (as noted above) for the field described by the given Lagrangian.

Suppose that quanta actually do interact in accordance with the direct action theory. Then there is no Fock space description of their interactions: such a space simply does not exist. In other words, the interaction picture (of quantized fields) really does not exist, just as Haag's theorem tells us. There is a certain interpretive elegance to this response to the theorem, analogous to abandoning the idea of the 'luminescent ether' in response to the negative result from the Michelson-Morley experiment.

But then, as noted by Earman and Fraser [2], the spectacular success of the interaction picture of QFT 'cries out' for explanation. The direct action theory of fields provides one: the 'quantized field' is not ontologically real but is rather a stand-in for the unknown, directly-interacting sources of the DAT. As shown by Narlikar, [8] (and Davies, [11-13] in the specific context of quantum electrodynamics), the direct action theory (DAT) is empirically equivalent to the quantized field picture (QFT), so QFT can be used as a calculational device for dealing with a reality actually described by the DAT. It is only when the stand-in entities (interacting field operators) are taken as fundamental that Haag's theorem becomes a threat. If instead the QFT picture is understood as an empirically equivalent but not fundamentally applicable model, then Haag's theorem simply tells us what we already know: the interaction picture of quantized fields does not really exist. What does exist is a non-quantized direct interaction that can be modeled, to empirical equivalence, by QFT.

---

[1] Actually, Wesley and Wheeler dissent from this common perception of direct-action theories as computationally cumbersome: "In addition to the conceptual simplicity of the theory, it is also more convenient mathematically. One need not calculate the dynamics of the field, a complex dynamical quantity with an infinite number of degrees of freedom; only the particles, with their finite number of degrees of freedom." [9], p. 428.

This author recognizes that direct-action theories are not currently popular, but given the severity of the threat posed by Haag's theorem, it may well be time to reconsider them. Indeed, one of the founders of the Wheeler-Feynman direct action theory of electromagnetism [10], the late John Wheeler, was recently doing just that in connection with the search for a theory of quantum gravity [9]. Together with D. Wesley in 2003, he reviews the history of the development of the Wheeler-Feynman (WF) theory and comments:

> [WF] swept the electromagnetic field from between the charged particles and replaced it with "half-retarded, half advanced direct interaction" between particle and particle. It was the high point of this work to show that the standard and well-tested force of reaction of radiation on an accelerated charge is accounted for as the sum of the direct actions on that charge by all the charges of any distant complete absorber. Such a formulation enforces global physical laws, and results in a quantitatively correct description of radiative phenomena, without assigning stress-energy to the electromagnetic field. ([9], p. 427)

Wesley and Wheeler note that one motivation for retaining the idea of a mediating field has been to enforce locality, and that some objections to the direct-action picture are based on an aversion to the idea of a 'nonlocal' interaction between particles; i.e., that the particles evidently interact instantaneously. They address this concern in a section entitled "Is Physics Entirely Local?," concluding that in fact it is not:

> One is reminded of an argument against quantum theory advanced by Einstein, Podolsky and Rosen in a well-known paper (1931) …The implicit nonlocality of [the EPR entanglement experiment], they argue, is at odds with the idea that physics should be fundamentally local…As has been evidenced by many experimental tests, the view of nature espoused by Einstein *et al* is not quite correct. Various experiments have shown that distant measurements can affect local phenomena. That is, *nature is not described by physical laws that are entirely local*. Effect from distant objects *can* influence local physics…this example from quantum theory serves to illustrate that it may be useful to expand our notions regarding what types of physical laws are 'allowed'. ([9], pp 426-7; emphasis in original text)

It should be clear from the above excerpts that the surviving original co-founder of the 'nonlocal' Wheeler-Feynman direct-action theory of electromagnetism views that formulation as perfectly viable. Moreover, he suggests that its nonlocal character should not be shunned but instead embraced, and that the same direct-action approach be applied towards longstanding stubborn challenges such as quantum gravity. In particular Wesley and Wheeler are questioning whether such challenges are fruitfully addressed by way of the usual conceptual tool of invoking a 'field' in order to try to account for the phenomena in a local manner. The present author would like to suggest that Haag's theorem is yet another challenge of this type, in which the 'local,' mediating field description has turned out to be fundamentally inadequate.

In perhaps a crude analogy, the mediating field plays the part of a 'bucket brigade' that is invoked in order to restrict the influence of the field to a local, continuous conveyance from spacetime point to spacetime point. (This is a key function of the commutation relations for the field operators, locality being enforced by suitable vanishing of the commutator.) But, as Wesley and Wheeler note, this sort of 'bucket brigade' account of physical influences apparently is not a feature of quantum entities. And the infinite number of degrees of freedom implicit in that local, mechanistic account clearly leads to various problems, such as Haag's theorem and the problem of infinite self-energies of field sources. These issues, as well as the advantage of a direct-action theory for the problem of gauge arbitrariness, are discussed further in the next section.

## 3. Various approaches to direct-action theories

A quantum relativistic version of the classical Wheeler-Feynman theory was developed in the early 1970s by Davies [11-13]. Davies noted that his theory naturally invokes the Coulomb gauge, since the Coulomb interaction is characterized by the time-symmetric propagator and can be considered a 'virtual photon' interaction only. In contrast, radiative phenomena in his theory correspond to Fock space states which must be on-shell and transversely polarized (i.e.. 'real photons'). (See [12], p. 843 for a discussion of the Coulomb gauge as the natural choice for the Davies direct-action theory.) An advantage of the Coulomb gauge is that it is a 'complete' gauge, i.e., lacking any residual arbitrariness, unlike other gauges such as the manifestly covariant Lorenz gauge.

A similar point, albeit arrived at from different perspective of seeking to avoid both the divergent energy of self-interaction and the 'light tight box' complete absorber condition, is made by F. Rohrlich:

> The solution to these difficulties came to me in the early sixties from the realization…that one wants to avoid only the self-interaction related to the Coulomb field and not the one related to radiation reaction…thus one is led to a theory which is of the action-at-a-distance type only for the Coulomb field but which remains a field theory with respect to the radiation field…
>
> This realization agrees beautifully with the quantum mechanical understanding of electromagnetic field: only the radiation field is composed of photons (i.e., must be quantized) while the Coulomb field is not (i.e., should not be quantized). This, in turn, leads evidently to the Coulomb gauge which is, in this sense, the natural gauge. In any case, the elimination of the Coulomb field is physically easily justified, the elimination of the radiation field, however, is not, because it would mean that the photon is not as elementary a particle as the electron, a notion that I find difficult to maintain on this level of theory. ([14], p. 350)

It should be noted that Rohrlich 's approach is a 'hybrid' one—he wishes to retain the quantized field for radiated photons but abolish it for the Coulomb interaction. One of his motivations was to eliminate the 'light tight box' boundary condition, and this can be done by using a quantized field for radiative processes only. However, the cost of this approach is arguably somewhat of a theoretical 'patchwork'.

Whatever approach to a direct action theory one wishes to pursue, the basic DAT picture evades Haag's theorem by denying that the interactions involve Fock space states. However, since we need a clear physical distinction between 'free' states and 'interacting' states to identify which entities are to be considered describable by Fock states and which are not, some matters of interpretation of DAT will be examined below.

The Wheeler-Feynman and Davies pictures form the theoretical basis of the transactional picture of quantum processes first developed by Cramer [15] and elaborated by the present author in a 'possibilist' and relativistic form [17-19] . In the latter, I have argued that the Davies theory naturally lends itself to a transactional account, in which radiative phenomena correspond to actualized transactions. The first step in a radiative process is the emission of a photon offer wave $|k>$ and confirming response from *all* accessible absorbers—even those that do not actually receive any real energy. Under PTI, the offer wave $|k>$ is identified as a true Fock space state, since it is on-shell and prompts an absorber response that makes possible the transfer of real energy via a transaction. This offer/confirmation exchange sets up a set of incipient transactions corresponding to momentum components $|\mathbf{k}>$ in all possible spatial directions, but (for a single photon) only one such direction can actually be chosen; that choice corresponds to the 'collapse' process.[2] This is the point at which one of the incipient transactions is actualized and a real photon is transferred (radiated) from an emitter to a particular 'winning' absorber. The transfer of a real photon with momentum **k** is represented by a projection operator $|\mathbf{k}\rangle\langle\mathbf{k}|$. Since the precursor to any such radiative process involves responses from all absorbers, the complete absorber response cannot be neglected.[3]

The above picture provides a unified explanation of the quantized radiation field in terms of actualized photons, even though the underlying dynamics is all mediated by direct-action. Because the radiated photons are quantized, PTI ends up being equivalent to Rohrlich's approach; but in PTI the quantization arises from the transactional process rather than being imposed by fiat.

The transactional picture also explains the apparently mysterious pole remaining in the Feynman propagator when it is derived, as in the Davies theory, from the confirming response of absorbers. Davies tacitly assumed that the Feynman propagator can remain applicable, at least in principle, as a description of virtual particle processes, since his primary aim was to demonstrate equivalence between the direct-action picture and standard QED. But in fact, as he shows by Fourier decomposition of the Feynman propagator into bound (time-symmetric, off-shell) and free (on-shell) parts (see [13], eq (5)), in the direct-action picture the internal lines in scattering diagrams are not really described by the Feynman propagator but rather by the time-symmetric propagator. Thus, the Feynman propagator becomes a 'hybrid' and somewhat awkward entity in the Davies account, since it is presumed capable of describing both virtual and real processes while imposing retarded propagation ('causality') on both. This ambiguity arises because neither Davies nor Feynman make a fundamental distinction between real and virtual photons at the level of the basic field propagation. In particular, Feynman considered them a matter of context.[4]

However, we need a clear physical distinction between the free and interacting field components in order to fully escape Haag's theorem, and PTI provides one. In PTI a confirming response from absorbers is identified as unambiguously leading to real photons, as opposed to virtual photons, and calls for the pole in the Feynman propagator, which corresponds to an on-shell, Fock state. The latter is an external line only in terms of a scattering diagram; it is not properly considered an internal line. True internal lines do not prompt an absorber response and that is why they can be accurately described by the time-symmetric component only [19], and why they are not correctly described by Fock states (which in the QFT interaction picture is what leads to the problem pointed to by Haag's theorem). As noted earlier, Rohrlich's picture, in which the Coulomb interaction is non-quantized and never transfers real energy, is very similar to the present author's 'possibilist' transactional account (PTI) in that virtual processes (i.e. the Coulomb interactions) do not rise to the level of incipient transactions, and therefore are not eligible to transfer conserved quantities such as energy and momentum via a real photon.

---

[2] This author has argued that the collapse can be understood as a kind of spontaneous symmetry breaking ([17], Chapter 4).
[3] As Davies notes in [13], p. 1035, when one does not include the full absorber response in the system under study, the direct-action theory involves a nonunitary scattering matrix. While Davies regards this as puzzling, in the transactional picture it is a natural reflection of the fact that full absorber response is a key part of any radiative process: radiated photons are always a product of the full absorber response, ultimately being absorbed by just one 'winning' absorber, and are not simply emitted as free-standing entities.
[4] Feynman has remarked that there is no fundamental difference between real and virtual particles ( [16], as quoted in Davies [13], pp. 1027-8 ). This is not the case in the transactional picture, as emphasized in Kastner [18], [19].

Again, the relevance of the treatment of virtual photons is that a key assumption required for Haag's theorem is that field states are defined for virtual processes. Indeed, one of the inelegant features of QFT is that the off-shell "states," formally subject to creation and destruction, must be eliminated by rampant use of Dirac delta functions as bookkeeping devices. In [1], pp 23-24, Haag notes that the delta function enforcing on-shell behavior "must appear in all relations of physical significance." (It would probably be more accurate to say 'empirical significance' in this connection, since the virtual photon exchanges are certainly physically significant. The problem is that they don't really correspond to states!)

The existence in QFT of creation and destruction operators for 'unphysical' states that must be eliminated in this *ad hoc* way points again to the fundamental problem: namely, the QFT model treats treats virtual propagation as physically equivalent to real propagation. However, the direct action theory makes a clear distinction between the two (at least as interpreted in PTI). In the latter, virtual processes are described by the time-symmetric propagator which does not correspond to a radiative process, and therefore does not correspond to a real photon or Fock space state. Thus, Haag's theorem is blocked by the direct-action approach.

An immediate additional side-benefit of the direct-action picture is gaining a physically natural basis for the choice of gauge, resolving another notorious problem of conceptual consistency and interpretation of relativistic field theories: apparent gauge arbitrariness.

**4. QFT evasions of Haag's theorem**

Earman and Fraser [2] observe that the conundrum presented by Haag's theorem can be circumvented in various ways within the QFT model, but it is generally acknowledged that these circumventions have their drawbacks and limitations. One such workaround is to ascribe to the interaction picture a 'renormalized' Hilbert space $\mathscr{H}_R$. $\mathscr{H}_R$ is the Hilbert Space on which the full Hamiltonian $H = H_F + H_I$ is defined, but not the free Hamiltonian $H_F$. Renormalization consists of introducing an infinite self-energy counterterm in the Hamiltonian—i.e., the divergence associated with the vacuum polarization energy is subtracted out. This addresses the immediate problem presented by the heuristic version of Haag's theorem by allowing the full Hamiltonian to annihilate the free field vacuum.

However, as noted by Earman and Fraser, this maneuver involves rejecting "the assumption that the ' + ' in $H_F + H_I$ should be taken to mean that each operator in the formal sum is separately well-defined on $\mathscr{H}_R$"; they add that "in fact only the combined operator has meaning." ([2, p. 315) But in fact each operator $H_F$ and $H_I$ *does* have meaning in the interaction picture; they are perfectly well-defined in terms of the free and interacting fields (eqs.1 and 2). Granted, the Hamiltonians' *actions on the states* are not well-defined, which is what Haag's theorem points to; nevertheless the fields themselves, of which the Hamiltonians are functionals, are physically well-defined in terms of their Lagrangians. Indeed the need in the QFT model to assume that field states exist for the interaction (which is denied in the direct-action picture) can again be seen as the source of the problem in this regard. Both approaches, QFT and DAT, use the same Lagrangians, but the direct-action picture avoids introducing the field states as intermediaries.[5]

Another workaround is Haag-Ruelle scattering theory [3], but this method only applies to massive particles, and therefore can treat massless particles only as an approximation. The constructive field theory approach of Glimm and Jaffe [22] is another approach attempting to surmount the problems brought to light by Haag's theorem, but this formulation has made only partial progress. At the end of his review [23], Jaffe presents a distinctly muted optimistic outlook by commenting that

---

[5] In this regard, it may be of (at least) historical interest to note that Feynman referred to renormalization as a 'shell game' and 'a dippy process' [20], although he seemed unaware at the time of Haag-type theorems. In more rigorous terms, Dirac [21] noted that renormalization involves "neglecting infinities which appear in [QFT's] equations, neglecting them in an arbitrary way. This is just not sensible mathematics. Sensible mathematics involves neglecting a quantity when it is small – not neglecting it just because it is infinitely great and you do not want it!"

> One can envision the positive future answer to the question of the existence of an asymptotically-free, four-dimensional gauge theory on a cylindrical space-time, although the infra-red (infinite-volume) limit still seems beyond grasp. (Jaffe 2000, p. 8)

Thus, it appear to this author that extant workarounds tend to be *ad hoc*, approximate, or partial measures and that the most sound approach in the face of Haag's theorem is to question the QFT model itself, rather than to try to retain the model by resorting to these kinds of modifications.

## 5. Conclusion

Teller has correctly observed that

> Everyone must agree that as a piece of mathematics Haag's theorem is a valid result that at least appears to call into question the mathematical foundation of interacting quantum field theory, and agree that at the same time the theory has proved astonishingly successful in application to experimental results. [7, p. 115]

If Nature is in fact described by a direct-action theory, then this apparent paradox is resolved: QFT is an empirically equivalent calculational stand-in for the direct-action theory, so it can continue to be used for practical calculations. Meanwhile, its mathematical inconsistencies can be rendered inconsequential since they can be understood as arising from its 'makeshift,' nonfundamental character. The other significant dividends gained by adopting the direct-action picture are: (i) a solution to the gauge arbitrariness problem and (ii) a solution to the self-energy problem of standard QED.

Acknowledgment
The author appreciates the valuable comments and constructive criticisms of an anonymous referee.